\documentclass[conference]{IEEEtran}

\usepackage{xspace}
\usepackage{balance}
\usepackage{graphicx}
\usepackage[hyphens]{url}
\usepackage{epstopdf}
\usepackage{microtype}
\usepackage{booktabs}
\usepackage[table,xcdraw]{xcolor}
\usepackage{caption}
\usepackage{subcaption}
\usepackage{amsmath}
\usepackage{enumitem}
\usepackage{amssymb}
\usepackage{pifont}
\newcommand{\ieee}{\mbox{IEEE~802.15.4}\xspace}

\newcommand{\NAME}{\mbox{\texttt{STX-Vote}}\xspace}

\newcommand{\fakepar}[1]{\smallbreak\noindent{}}
\newcommand{\boldpar}[1]{\smallbreak\noindent\textbf{#1.}}

\iffalse
 \newcommand{\comment}[1]{\textcolor{blue}{#1}\xspace}
 \else
 \newcommand{\comment}[1]{#1}
 \fi

\usepackage{manyfoot}
\DeclareNewFootnote{A}
\DeclareNewFootnote{B}

\let\footnoteR\footnoteB
\let\footnote\footnoteA
\iftrue
    \newcommand{\br}[1]{\footnoteR{{\color{blue}\bf BR: #1}\color{blue}}}
    \newcommand{\mb}[1]{\footnoteR{{\color{red}\bf MB: #1}\color{red}}}
    \newcommand{\ih}[1]{\footnoteR{{\color{red}\bf IH: #1}\color{red}}}
\else
    \newcommand{\br}[1]{}
    \newcommand{\mb}[1]{}
    \newcommand{\ih}[1]{}
\fi
\iffalse
\author{
\IEEEauthorblockN{Anonymous Authors} 
}
\else
\author{
\IEEEauthorblockN{
Burhanuddin Rangwala\IEEEauthorrefmark{1}, Ava Powelson\IEEEauthorrefmark{1},
Michael Baddeley\IEEEauthorrefmark{2}, 
and 
Israat Haque\IEEEauthorrefmark{1}} 
\IEEEauthorblockA{\IEEEauthorrefmark{1}Dalhousie University, Canada, \IEEEauthorrefmark{2}Technology Innovation Institute, UAE }
}\vspace{-3.50mm}
\fi

\begin{document}

\title{\NAME: Improving Reliability with Bit Voting in Synchronous Transmission-based IoT Networks \vspace{-3.00mm}}

\maketitle



\begin{abstract}

Industrial Internet of Things (IIoT) networks must meet strict reliability, latency, and low energy consumption requirements. However, traditional low-power wireless protocols are ineffective in finding a sweet spot for balancing these performance metrics. Recently, network flooding protocols based on Synchronous Transmissions (STX) have been proposed for better performance in reliability-critical IIoT, where \textit{simultaneous} transmissions are possible without packet collisions. STX-based protocols can offer a competitive edge over routing-based protocols, particularly in dependability. However, they notably suffer from the beating effect, a physical layer phenomenon that results in sinusoidal interference across a packet and, consequently, packet loss. Thus, we introduce \NAME, an error correction scheme that can handle errors caused by beating effects. Importantly, we utilize transmission redundancy already inherent within STX protocols so \textit{do not incur additional on-air overhead}. Through simulation, we demonstrate \NAME can provide a 40\% increase in reliability. We subsequently implement \NAME on \texttt{nRF52840-DK} devices and perform extensive experiments. The results confirm that \NAME improves reliability by 25-28\% for BLE\,5 PHYs and 8\% for IEEE 802.15.4; thus, it can complement existing error correction schemes.


\end{abstract}

\section{Introduction}
\label{sec:introduction}

Traditionally, wireless network communication protocols, such as Carrier Sense Multiple Access (CSMA) or Time Division Multiple Access (TDMA), have been designed assuming that packet collisions are inherently destructive and should be avoided as much as possible. Methods such as carrier sensing, handshaking, and transmission scheduling have been utilized to avoid such collisions. However, after Ferrari et al.~\cite{glossy} showed that by tightly synchronizing transmissions, the packet collisions were not inherently damaging in certain physical-layer standards, there has been a considerable body of research into network flooding protocols based on Synchronous Transmissions (STX)~\cite{zimmerling20synchronous} (also referred to in literature as \textit{Concurrent Transmissions}). Such protocols can significantly reduce latency and power consumption as various complexities and overhead of MAC and routing can be eliminated. 


In theory, given perfect temporal and frequency synchronization between transmitters, signals overlap constructively and increase the probability of a packet's successful reception at its destination. However, in real-world scenarios, the carrier frequency of the radios of transmitting devices are typically offset due to imperfections in their crystal oscillators~\cite{baddeley2023understanding}. This relative Carrier Frequency Offset (CFO) between devices causes \emph{beating}, a sinusoidal waveform of \textit{both} constructive and destructive interference. The destructive interference period in beating weakens the signal and can create errors in transmissions, negatively affecting reliability.
%
%

STX-based networking protocols typically use re-transmissions~\cite{glossy,RedFixHop,rof}, channel hopping~\cite{rof,harmony}, network coding~\cite{harmony,Mixer,codecast}, or a combination of these, to increase reliability. These methods exploit the \textit{constructive interference} or \textit{capture effect}, which is susceptible to packet loss when the number of transmitters increases. There are methods such as Forward Error-Correction (FEC) or interleaving at the physical layer, but their impact on reliability in STX protocols is yet to be explored~\cite{zimmerling20synchronous}. Baddeley et al.~\cite{baddeley2023understanding} point out that errors occurring due to beating can be discreet or burst depending on network conditions. However, FEC or Direct Sequence Spread Spectrum (DSSS) cannot handle burst errors where contiguous bits are corrupted beyond a certain point, as indicated in~\cite{liao2016revisiting}. Consequently, coded BLE\,5 physical layers and \ieee are vulnerable to decoding errors under certain beating conditions.
%
%


Though the authors of~\cite{baddeley2023understanding} show that protocols would benefit from an error-correction mechanism capable of handling both burst and discrete errors by harnessing repetitions in STX protocols, no such solution has yet been proposed. This paper addresses this gap and proposes \NAME, a novel packet-level error correction scheme that \textit{leverages the multiple repetitions inherent in commonly used STX protocols} to combat beating-based packet loss and improve reliability. As such, \NAME also \textit{does not incur additional transmission overhead}. In the event of an error, \NAME attempts to create a correct packet out of previously received error packets. Unlike traditional error correction schemes, \NAME is specifically targeted at STX-based protocols, where errors due to beating are likely to be different for each (re)transmission of a particular packet.


\boldpar{Our contributions} By extending an open-source STX simulation~\cite{baddeley2023understanding}, we firstly explore how bit-voting can improve reliability across BLE\,5 and IEEE~802.15.4 physical layers. We subsequently implement \NAME on the \texttt{nRF52840-DK} platform within the Open\,Synchronous\,Flooding\,(OSF) framework~\cite{baddeley2022osf} and conduct an extensive evaluation in a local and the D-Cube public testbed~\cite{dcube1}. We compare STX reliability both with and without \NAME across the nRF's BLE\,5 and IEEE~802.15.4 physical layers. We show that using \NAME under strong wide and narrow beating conditions increases reliability by 25-28\% and 8\% on the BLE\,5 and IEEE~802.15.4 PHYs, respectively. \NAME is available as open source\footnote{\url{https://github.com/PINetDalhousie/STX_Vote}} for reproducibility and extension. 

\boldpar{Paper outline} We provide detailed background information on the beating effect in Sect.~\ref{sec:background}, and explore the wider issue of reliability in STX protocols in Sect.~\ref{sec:related-work}. In Sect.~\ref{sec:design}, we cover design and implementation details on the bit-voting technique behind \NAME, while simulation and experimental results are provided in Sect.~\ref{sec:evaluation}. Finally, we conclude the paper in Sect.~\ref{sec:conclusion}.

\section{Background and Motivation}
\label{sec:background}
\begin{figure}[t!]
    \centering
    \includegraphics[width=0.8\columnwidth]{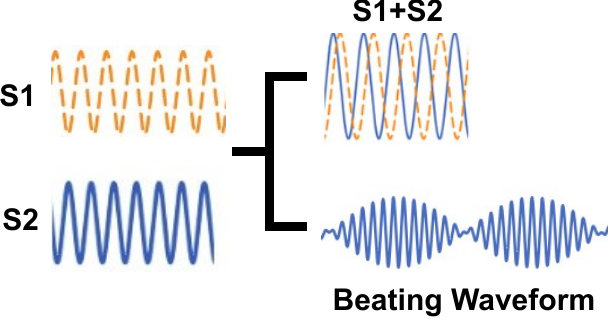}
    \vspace{-2.00mm}
    \caption{Beating effect when frequency offset transmissions overlap.}
    \label{fig*:beating}
\end{figure}
\vspace{-2.00mm}
\begin{figure}[t!]
    \centering
    
    \includegraphics[width=1.0\columnwidth]{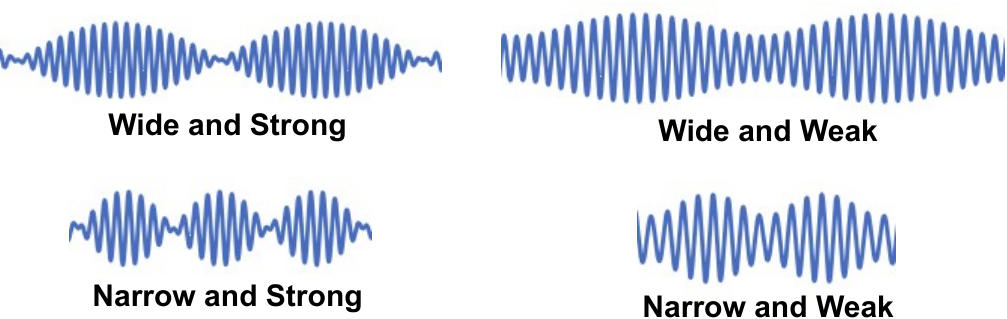}
    \vspace{-5.00mm}
    \caption{Types of beating~\cite{baddeley2023understanding}.}
    \label{fig*:beating_types}
\end{figure}
\vspace{2.00mm}
The \textit{beating effect} is a phenomenon that occurs when signals from non-coherent transmitters with slightly different carrier frequencies overlap in the air~\cite{baddeley2023understanding}. This results in a sinusoidal waveform of alternating periods of constructive \textit{and} destructive interference. Although devices transmit at the same carrier frequency, imperfections in their crystal oscillators lead carrier frequencies to deviate from the intended level and give rise to a relative carrier frequency offset (CFO) between the transmitted waveforms (Fig.~\ref{fig*:beating}). Bits falling in destructive valleys tend to get corrupted, causing errors in the packet. Baddeley~et~al.~\cite{baddeley2023understanding} showed that the relative CFO between devices, the time for which the packet is in the wireless transmitting medium, and the power difference between synchronous transmissions give rise to four basic types of beating patterns as shown in Fig.~\ref{fig*:beating_types}.\\

\begin{itemize}
    \item \textbf{Wide and Strong:} Occurs when the CFO between the devices and the power difference between signals is small. Depending on the packet air time, destructive valleys in the waveform can last longer during packet transmission, possibly resulting in burst errors.
    \item \textbf{Wide and Weak:}  Occurs when the CFO between the devices is small, and the power difference between signals is significant but less than the capture threshold. The destructive valleys in the waveform are not as intense as wide and strong, and will generate burst errors. Power differences larger than the capture threshold may lead to the capture effect. 
    \item \textbf{Narrow and Strong:} Occurs when the CFO between devices is large, and the power difference between signals is small. The narrow destructive valleys in the beating waveform are more likely to cause discrete errors across the packet.
    \item \textbf{Narrow and Weak:} Occurs when the CFO between devices is small, and the power difference between signals is significant. The destructive valleys in the beating waveform are less intense compared to narrow and strong beating and will result in discrete errors. 
\end{itemize}


Importantly, different physical layers behave differently depending on how the beating manifests. Extensive experimentation in~\cite{baddeley2023understanding} concludes that the uncoded BLE\,5 PHYs can perform better in \textit{wide} beating scenarios due to the fact that fast transmission rates and multiple re-transmissions increase the probability of one of the transmissions occurring during a constructive beating peak. However, they suffer during \textit{narrow} beating as the constructive beating peaks are narrow, and they do not have any error correction mechanism that can help mitigate narrow beating.

On the other hand, coded BLE\,5 PHYs perform better during \textit{narrow} beating because the forward error-correction (FEC) mechanism used in BLE\,5 coded PHYs (125K, 500K) and the Direct Sequence Spread Spectrum (DSSS) used in IEEE~802.15.4 can correct burst errors up to a certain extent~\cite{baddeley2023understanding}. For example, in the case of IEEE~802.15.4, if the width of the error is narrower than the DSSS symbol length, then correction is possible~\cite{liao2016revisiting}. To handle burst errors caused by wide beating using these correction mechanisms, one might suggest increasing the redundancy in FEC or the DSSS symbol length. However, the number of bits affected by beating cannot be quantified as it occurs randomly. Also, the error handling steps can increase the decoding time at the receiver and the coding overhead at the sender.
Finally, the redundancy added by convolution codes in the coded BLE\,5 PHYs might get corrupted in the packets themselves, leading to unsuccessful error correction attempts.

Capture effect, where a signal of significantly higher power than other signals can be correctly demodulated, is another way to mitigate beating, as suggested in~\cite{baddeley2023understanding}. However, for the capture effect to work, the power of the dominant transmitter must be higher than the sum of other simultaneous transmissions for a packet reception to be possible at the receiver~\cite{escobarthesis}, which can negatively affect the scalability and simplicity advantages of using STX-based protocols.

Overall, STX-based protocols are complex and cannot scale to mitigate the beating effect~\cite{baddeley2019competition}. Additionally, FEC and DSSS can add communication overhead that can lead to synchronization drift, thus degrading the reliability. The proposed \NAME fills the above gaps by introducing an error correction mechanism that does not add overhead while successfully mitigating the beating effect both in discrete and burst errors. Also, \NAME can complement the existing solutions and improve their reliability.
\newpage
\section{Related Work}
\label{sec:related-work}

\boldpar{Channel Hopping} There are several works that use channel hopping as a method to increase network resilience. Lim et al.~\cite{rof} employed channel hopping for every communication slot to avoid any possible interference that a previous transmission slot might have encountered. The authors of ~\cite{inter_rs_ch} focused on improving resiliency of the CRYSTAL protocol using channel hopping and noise detection. The Transmission and Acknowledgment (TA) pairs of the protocol transmit on different channels to overcome the limitation of the longer awake time of the noise detection method. RedFixHop~\cite{RedFixHop} is another protocol that uses channel hopping. Every node in all of these protocols either performs channel assessment for adaptive channel-hopping or determines a fixed channel-hopping sequence among nodes. 
Beating can still occur under channel hopping because of the Relative Frequency Offset (RFO) between the transmitting devices, leading to packet corruption. Thus, these solutions need an error correction mechanism while considering the beating effect.
%
\boldpar{Network Coding} Network coding employs mathematical algorithms to combine multiple message units from different nodes into a single transmission towards the receiver. In STX-based networks,it leverages the capture effect to enhance reception probability and reliability by giving multiple possibilities for the message encoded in different simultaneous transmissions to reach the destination. Splash~\cite{splash} uses XOR coding, Ripple~\cite{ripple} uses Reed-Solomon coding focused on one-to-many communication, while Mixer~\cite{Mixer} uses Random Linear Network coding (RLNC) targeting many-to-many communication. However, these solutions do not consider the beating effect and can suffer from the same limitations as in channel hopping. For more works on STX, the reader is directed to~\cite{zimmerling20synchronous}.

\boldpar{Re-transmissions} Ferrari et al.~\cite{glossy} (using \textit{Rx-Tx...} pattern) and RobustFlooding~\cite{rof} (using \textit{Rx-Tx-Tx...} pattern) use aggressive re-transmission of packets to exploit the possibility that at least one of the transmissions will reach the destination. All nodes using these communication protocols listen to all packets being transmitted in the network and follow their transmission patterns, which increases the probability of successful reception using either constructive interference or the capture effect. Because of the beating effect, if one of the nodes does not receive the correct transmission in a slot, then it may suffer from synchronization drift and impact the entire flooding. Thus, we need an error correction scheme. 

\boldpar{Packet Combining} This error correction mechanism relies on incorrect packet receptions: when there are enough received error packets, packet merging is performed to find the correct packet. The authors of~\cite{dubois2005packet} perform XOR operations on the incorrect packets to find the error positions and then employ a brute-force method to predict the bit value and use a CRC to check which bit values make up the correct packet. Received Signal Strength Indicator (RSSI) based packet combining~\cite{sakdejayont2016beating} uses the symbol with the highest power among the received incorrect packets to form a new packet and uses error checks to verify that the packet created is correct and then transmits it. Finally, the authors of~\cite{kothapalli2022extended} make a database of corrected packets based on the error positions and then use it to estimate the possible correct packet.  
The packet combining with FEC is useful to mitigate the beating effects due to its periodic nature~\cite{dubois2005packet}. However, packet combining is not deployed in synchronous flooding to tackle errors generated due to beating. This work fills that gap and proposes \NAME.    



\section{Error Correction through Bit-Voting}
\label{sec:design}

As introduced in Sect.\,\ref{sec:background}, STX-based communications can suffer from transmission errors (beating errors) due to the discrepancy between oscillators of a communicating pair. However, existing BLE\,5 and IEEE~802.15.4 physical layers cannot handle such errors at receiver.~\cite{baddeley2023understanding}. 

\begin{figure}[ht]
    \centering
    \includegraphics[width=1.0\columnwidth]{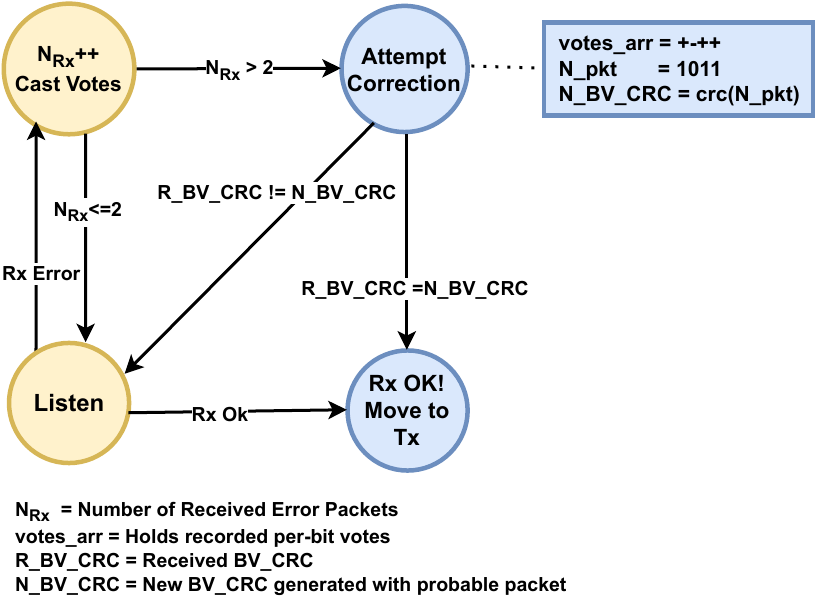}
    \caption{\NAME workflow at a receiver.}
    \label{fig:voting_det}
\end{figure}
The phase of the beats occurring can be different in different receptions of the same packet; thus, a specific contiguous block of bits may be corrupted in one reception but be received correctly in another. Based on this observation, \NAME votes on the values of the received bits. If the majority voting happens on the correct portion of a packet, then it can successfully correct the packet. Specifically, \NAME works at the receiver, as shown in Fig.~\ref{fig:voting_det}, by maintaining an array of signed integers that represents votes on the value of each bit in received error packets, where these error packets contain the same payload across multiple communication rounds. This voting array is used to construct the intended packet and validate it against a special CRC appended by the sender. 


\subsection{Voting Process}\label{sec:voting}

A reception is considered incorrect if the CRC check, \comment{performed by the receiver's hardware, fails}. The receiver maintains votes on the value of each bit among multiple incorrect receptions of the same packet. The voting array is reset every time a new unique packet is received (including a correct one).  

\begin{figure}[ht]
    \centering
    \includegraphics[width=.9\columnwidth]{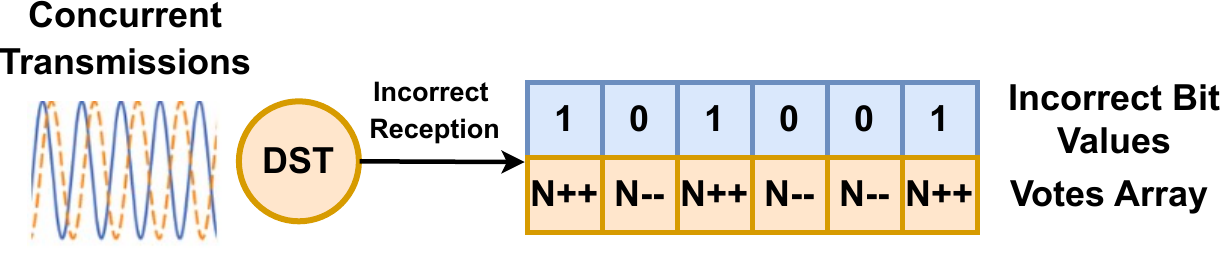}
    \caption{The voting operation of \NAME at a receiver.}
    \label{fig:voting_fig}
    \vspace{-6.00mm}
\end{figure}

As shown in Fig.~\ref{fig:voting_fig}, if the bit value at index zero in a received error packet is 1, the integer value at index zero in the voting array is incremented. The integer value is decremented if the bit value is 0. Thus, if a node receives several error packets with the same payload in a row, the votes will accumulate and determine each bit’s likeliness of being 1 or 0 based on the sign of the integer, which is illustrated in Section~\ref{correction}.

\subsection{Error Correction}\label{correction}

When a ROF (Robust Flooding)~\cite{rof} communication round ends without correct reception, a receiver performs the error correction. Specifically, the sender calculates a 2-byte CRC on each packet and appends it at the end. If the receiver receives two or more incorrect packets, it uses the corresponding voting array to construct the correct packet and its CRC. In particular, if the voted value at an index is positive, the corresponding bit value of the packet is set to 1; otherwise, it is 0.

As shown in Fig.~\ref{fig:voting_det}, a CRC is calculated from this constructed packet (ignoring the last 2 bytes of the packet, which is the constructed CRC). If this calculated CRC is identical to the constructed CRC, the packet reception is considered successful based on successful packet recovery. Otherwise, the packet is not recovered, and the receiver waits for the following receptions, which can lead to further voting on the packet or receiving a correct reception. In the case of correct reception, all error correction activities are suspended, i.e., the receiver does not perform error correction. Otherwise, the receiver initiates voting and follows the same correction procedure at the end of the round in the case of errors. Voting arrays and relevant counters are reset based on the packet ID for every new packet. 
\section{Evaluation}
\label{sec:evaluation}

We first assess the performance of STX over IEEE\,802.15.4 and BLE physical layers and measure the corresponding Packet Error Rate (PER) in a simulator to highlight the effect of different types of beating. Next, we show how \NAME can reduce the beating errors in STX-based communications. Finally, we conduct a series of experiments measuring the Packet Delivery Ratio (PDR) on real platforms to show the efficiency and deployment feasibility of \NAME in complementing existing error correction mechanisms. 

\subsection {Simulation}
\begin{figure}[!ht]
    \centering
    \begin{subfigure}[ht]{0.23\textwidth}
        \centering
        \includegraphics[width=\columnwidth]{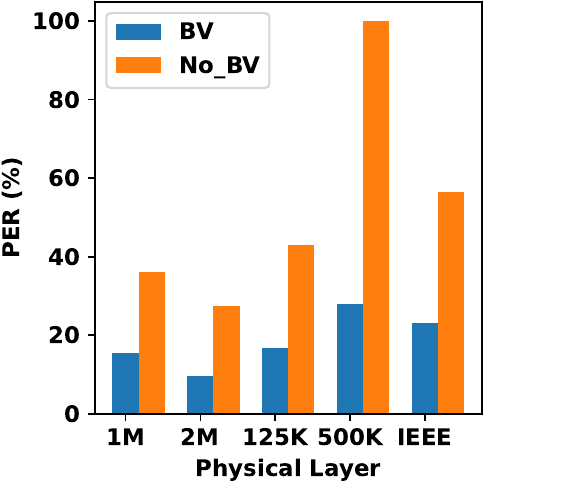}
        \caption{Wide~(RFO = 500Hz - 2000Hz)}
        \label{fig:simulation_swide_pic}
    \end{subfigure}
    \begin{subfigure}[ht]{0.23\textwidth}
        \centering
        \includegraphics[width=\columnwidth]{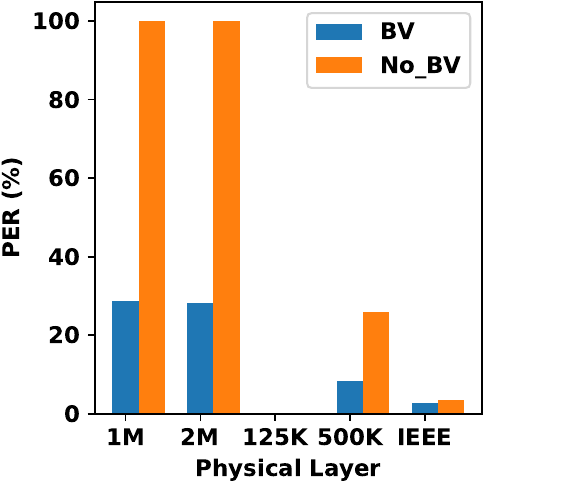}
    \caption{Narrow~(RFO = 10KHz - 40KHz)}
    \label{fig:simulation_snarrow_pic}
    \end{subfigure}
    \vspace{1mm}
    \begin{subfigure}[ht]{0.23\textwidth}
        \centering
        \includegraphics[width=\columnwidth]{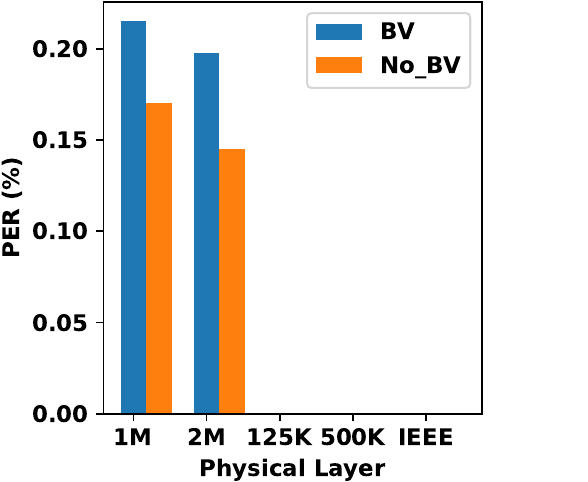}
        \caption{Wide~(RFO = 500Hz - 2000Hz)}
        \label{fig:simulation_wwide_pic}
    \end{subfigure}
    \begin{subfigure}[ht]{0.23\textwidth}
        \centering
        \includegraphics[width=\columnwidth]{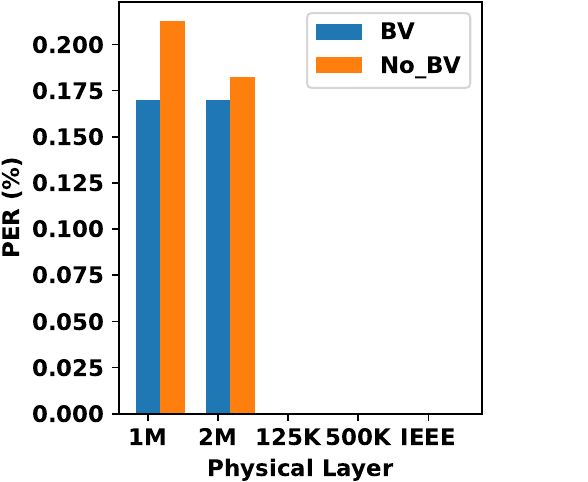}
    \caption{Narrow~(RFO = 10KHz - 40KHz)}
    \label{fig:simulation_wnarrow_pic}
    \end{subfigure}
    \caption{The average PER for strong and weak beatings.}
    \vspace{-4.00mm}
    \label{fig:Simulation_Pics}
\end{figure}

\textbf{Setup.}
We adapt the simulator\footnote{\url{https://github.com/ADEscobar/ct-simulator}}~\cite{baddeley2023understanding} to evaluate the packet error rate of \NAME with two synchronous transmitters over the four BLE and the IEEE\,802.15.4 PHYs across different beating frequencies. We assume an additive white Gaussian noise and no synchronization errors while simulating different CFOs between two transmitters. Different CFOs result in wide and narrow beating frequencies, while the transmission power decides weak and strong beating. We generate a random packet stream with eight samples per symbol following~\cite{escobarthesis}, where two synchronous transmissions suffer from amplitude distortion if their identical transmissions overlap in the air. Finally, we implement the proposed bit voting, error detection, and error correction mechanism of \NAME with a validation process by comparing the transmitted packets with the reconstructed ones.

\textbf{Discussion on results.} 
Fig.~\ref{fig:Simulation_Pics} shows the average PER over different beating frequencies showcasing wide and narrow beating with a 25dBm Signal-to-Noise ratio. These beatings can be strong or weak. We will first present the results for the strong one, which is shown in Fig.~\ref{fig:simulation_swide_pic} and Fig.~\ref{fig:simulation_snarrow_pic}. Uncoded PHYs (BLE\,1M and BLE\,2M) offer an increase in PER as the beating frequency increases. In the case of narrow beating, PER increases as seen in Fig.~\ref{fig:simulation_snarrow_pic} as packet transmission spans one or more destructive valleys, resulting in frequent errors. Uncoded PHYs cannot handle these errors due to the lack of an error correction mechanism despite having high transmission rates. However, we observe a significant decrease in PER from 67.88\% to 21.39\% for BLE\,1M and from 63.57\% to 18.78\% for BLE\,2M  when \NAME is enabled, confirming its effectiveness over strong narrow beating. In the case of wide beating, i.e., when an RFO between synchronous transmitters is not high, the PER is not as high as compared to narrow beating as seen in Fig.~\ref{fig:simulation_swide_pic} because their fast transmission rates increase the probability of one of the transmissions occurring during a constructive beating peak, however they require \NAME for the transmissions lying in the beating valley while experiencing strong wide beating which is why we can see a 15\% decrease in PER for BLE\,1M and 18\% decrease in PER for BLE\,2M.

Coded PHYs (BLE\,125K, BLE\,500K and IEEE\,802.15.4) are equipped with error correction mechanisms. For instance, FEC for BLE\,125K and BLE\,500K, while IEEE802.15.4 uses DSSS. The role of these mechanisms under beating is evident from Fig.~\ref{fig:simulation_snarrow_pic}. These mechanisms can handle narrow beating, giving similar performance to \NAME, which suggests that the coding used by BLE\,125K, BLE\,500K and IEEE802.15.4 can effectively correct the shorter error bursts introduced in the packet due to narrow beating. However, the trend is reversed in wide beatings as seen in Fig.~\ref{fig:simulation_swide_pic}, which may stem from their distortions that spread across multiple symbols and require a robust error correction mechanism like \NAME. It reduces the errors by 15\% for BLE\,125K, by 23\% for IEEE\,802.15.4 and by 63\% for BLE\,500K. 

Also, we conclude that weak beating does not significantly impact PER, as seen in Fig.~\ref{fig:simulation_wwide_pic} and Fig.~\ref{fig:simulation_wnarrow_pic}. In that case, signal distortion due to the higher transmission power of a signal compared to other concurrent ones is not strong enough to corrupt a bit. Thus, uncoded PHYs do not require any error correction mechanism in the presence of weak beating. Using their existing error correction mechanisms, coded PHYs can handle such signal distortion due to weak beating. 

\subsection{Lab Experiments}

\textbf{Setup.}  We use 3 \texttt{nRF52840-DK} devices that support all BLE PHYs and IEEE802.15.4 and place them on a desk at a random distance to create a single-hop communication among the nodes, assuming no synchronization errors. These nodes perform specific tasks according to the assigned roles, e.g., a source generates packets of length 255 bytes for BLE and 125 bytes for \ieee, a forwarder (also acts as another synchronous transmitter), and a destination node. We use a transmission power of -40dBm. The chosen parameters allow us to observe beating manifestation and identify the types of beating occurring. The source generates 100 packets for each PHY to measure the average Packet Delivery Ratio. 


\begin{figure*}[!ht]
    \centering
        \begin{subfigure}[ht]{0.19\textwidth}
        \centering
        \includegraphics[width=\columnwidth]{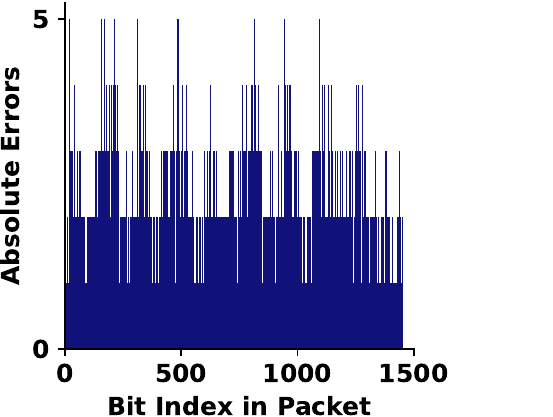}
        \caption{1M}
        \label{fig:beating_lcl_1m_pic1}
    \end{subfigure}
    \begin{subfigure}[ht]{0.19\textwidth}
        \centering
        \includegraphics[width=\columnwidth]{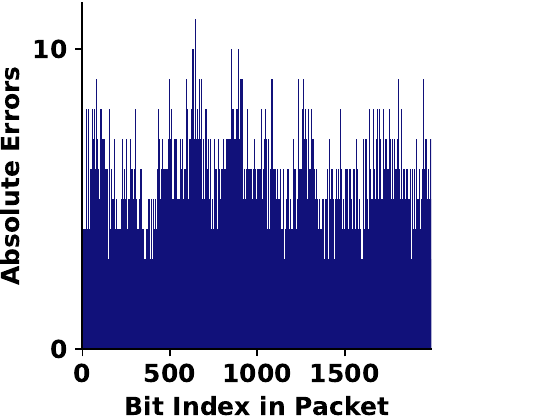}
        \caption{2M}
        \label{fig:beating_lcl_2m_pic1}
    \end{subfigure}
    \begin{subfigure}[ht]{0.19\textwidth}
        \centering
        \includegraphics[width=\columnwidth]{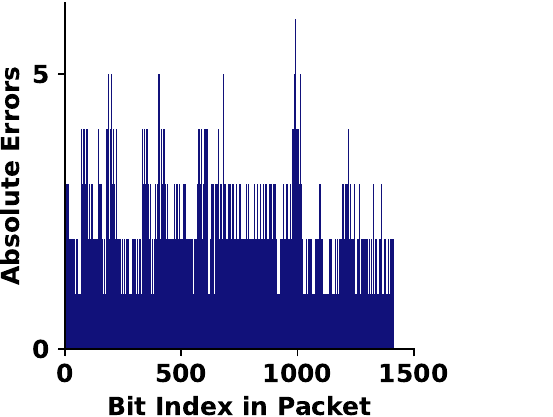}
        \caption{125K}
        \label{fig:beating_lcl_125K_pic1}
    \end{subfigure}
    \begin{subfigure}[ht]{0.19\textwidth}
        \centering
        \includegraphics[width=\columnwidth]{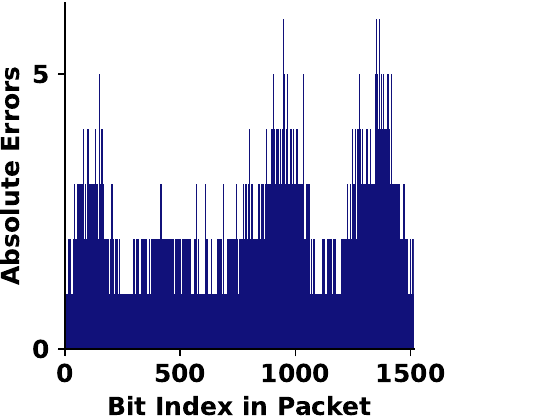}
        \caption{500K}
        \label{fig:beating_lcl_500K_pic1}
    \end{subfigure}
    \begin{subfigure}[ht]{0.19\textwidth}
        \centering
        \includegraphics[width=\columnwidth]{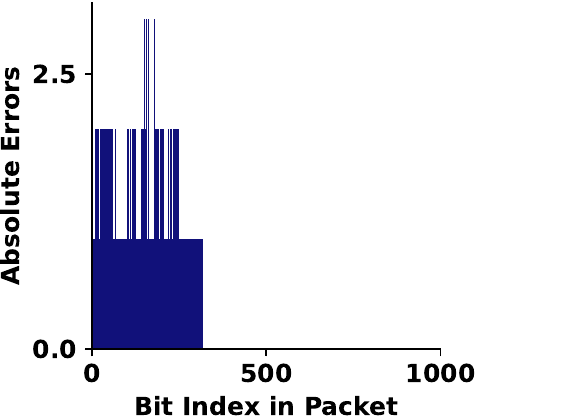}
        \caption{IEEE}
        \label{fig:beating_lcl_ieee_pic1}
    \end{subfigure}
    \vspace{-2.00mm}
    \caption{Beating patterns in the physical layers over local experiments. }
    \label{fig:local_beating}
    \vspace{-2.00mm}
\end{figure*}
\begin{figure*} [!ht]
    \centering
    \begin{subfigure}[ht]{0.19\textwidth}
        \centering
        \includegraphics[width=\columnwidth]{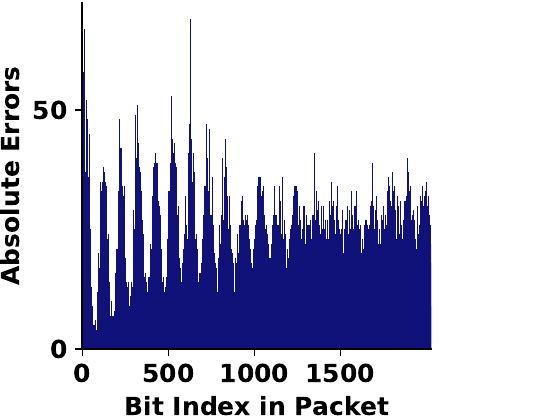}
        \caption{1M}
        \label{fig:beating_n20_1m_pic}
    \end{subfigure}
    \begin{subfigure}[ht]{0.19\textwidth}
        \centering
        \includegraphics[width=\columnwidth]{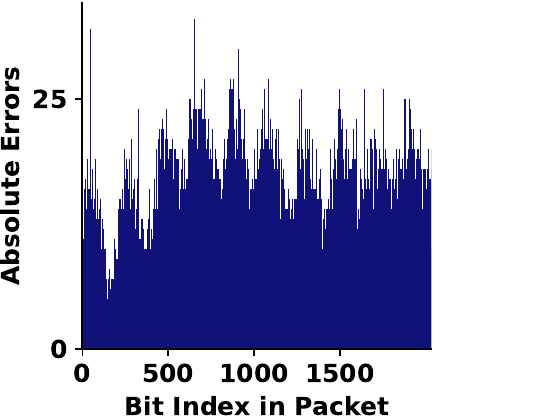}
        \caption{2M}
        \label{fig:beating_n20_2m_pic}
    \end{subfigure}
    \begin{subfigure}[ht]{0.19\textwidth}
        \centering
        \includegraphics[width=\columnwidth]{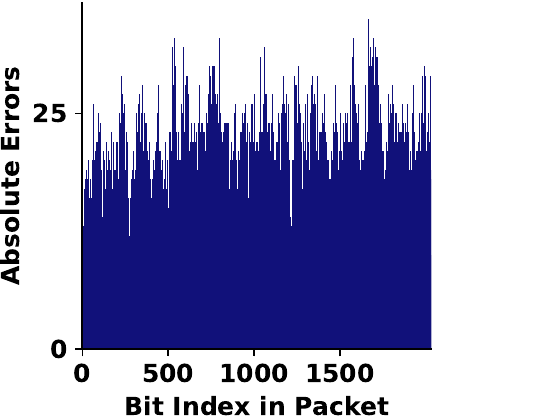}
        \caption{125K}
        \label{fig:beating_n20_125K_pic}
    \end{subfigure}
    \begin{subfigure}[ht]{0.19\textwidth}
        \centering
        \includegraphics[width=\columnwidth]{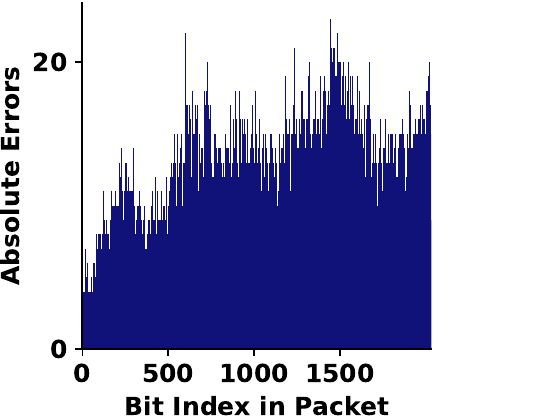}
        \caption{500K}
        \label{fig:beating_n20_500K_pic}
    \end{subfigure}
    \begin{subfigure}[ht]{0.19\textwidth}
        \centering
        \includegraphics[width=\columnwidth]{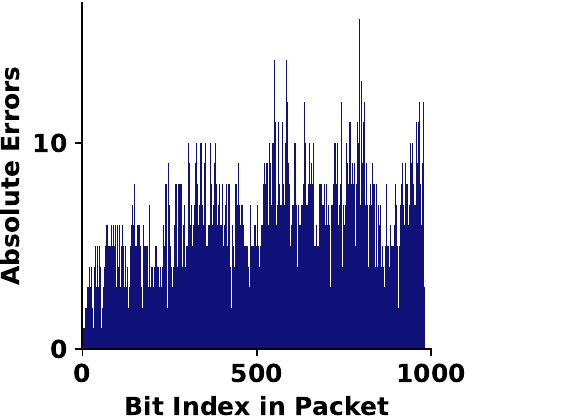}
        \caption{IEEE}
        \label{fig:beating_n20_ieee_pic}
    \end{subfigure}
    \vspace{-2.00mm}
    \caption{Beating encountered in D-cube experiments.}
    \label{fig:beating_pic_1_n20}
    \vspace{-5.00mm}
\end{figure*}

\begin{figure}[ht]
    \centering
    \begin{subfigure}[ht]{0.49\columnwidth}
        \vspace{5.00mm}
        \includegraphics[width=\columnwidth]{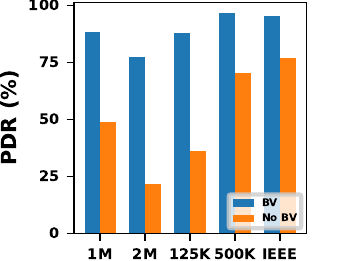}
        \caption{Local.}
        \label{fig:plt_avg_pdr_ng40_lcl}
    \end{subfigure}
    \begin{subfigure}[ht]{0.49\columnwidth}
        \vspace{5.00mm}
        \includegraphics[width=\columnwidth]{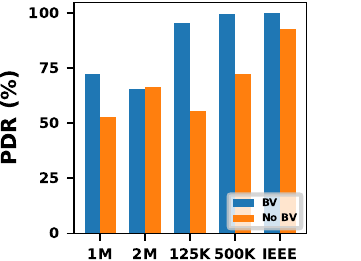}
        \caption{D-Cube.}
        \label{fig:plt_avg_pdr_neg20}
    \end{subfigure}
    \caption{The average PDR for local and D-cube experiments.}
\end{figure}

\textbf{Discussion on results.} Fig.~\ref{fig:plt_avg_pdr_ng40_lcl} presents the average PDR with and without using \NAME. We observe that \NAME improves the overall PDR for all PHYs. For instance, in uncoded PHYs, BLE\,1M and BLE\,2M obtained a performance gain of approximately 40\%  and 55\%, respectively. BLE\,2M suffers from the beating effect the most due to its high transmission rate, the lack of an error correction mechanism, and the presence of narrow beating as seen in Figs.~\ref{fig:beating_lcl_1m_pic1}~and~\ref{fig:beating_lcl_2m_pic1}. \NAME successfully detects and corrects its corrupted packets at receivers. Specifically, due to its high transmission rates and narrow beating, BLE\,2M suffers from discrete errors that corrupt a small portion of a packet. Also, due to the periodic nature of the beats, this corruption manifests in different parts of a corrupted packet for each transmission. Thus, voting on multiple such packets results in successful corrections in BLE\,2M. Similarly, BLE\,1M gains performance by adopting \NAME, where the degree of improvement is proportional to its transmission rates.  


On the other hand, coded PHYs BLE\,125K, BLE\,500K, and IEEE\,802.15.4 offer approximately 50\%, 20\%, and 20\%, improvements respectively with \NAME enabled. These PHYs suffer from burst errors generated by wide beating (Figs.~\ref{fig:beating_lcl_125K_pic1}~to~\ref{fig:beating_lcl_ieee_pic1}) with BLE 125K being the worst one. This behaviour is due to its lowest transmission rate and ineffective error correction using FEC. However, incorporating \NAME BLE 125K can successfully correct the receptions that its underlying FEC misses. Similarly, \NAME complements the FEC of BLE 500K and DSSS of IEEE\,802.15.4 with a performance improvement. All the PHYs can increase their PDR, but uncoded PHYs can significantly increase their PDR compared to coded PHYs using \NAME.

\vspace{-2.00mm}
\subsection{Experiment on D-cube Testbed }

\textbf{Setup.} Over-the-air experiments were performed on the D-cube public testbed~\cite{dcube1}, which is equipped with \texttt{nRF52840-DK} devices that support all BLE PHYs and IEEE~802.15.4. Three single-hop nodes were used in the experiments: a source node that generated and sent packets with randomized payloads of 246 Bytes when \NAME was enabled (so that the CRC generated by \NAME and packet header could be added, resulting in the packet size of 255 bytes), a destination node, and a forwarder node that acted as another synchronous transmitter.


Since multiple node combinations are possible in D-cube, we conducted an exploratory experimentation drive to find the node combinations that showed beating that was destructive enough for the physical layers to encounter errors from which their respective error correction schemes could not recover. Thus, \NAME could correct errors and increase the reliability of the network.

For this experimentation drive, we selected ten random three-node combinations with roles as used in the previous experimentation and ran tests using the transmission powers -16dBm, -20dBm, and -40dBm. In each case, the source transmitted 200 packets of length 255 bytes for the BLE physical layers and of size 128 bytes for IEEE\,802.15.4. We present the results for -20dBm due to the space constraints (the trend is similar in other cases). 


\textbf{Discussion on results.}
Fig.~\ref{fig:plt_avg_pdr_neg20} shows a significant improvement of almost 20\% in PDR for BLE\,1M, but no improvement for BLE\,2M. The beating pattern of 1M is shown in Fig.~\ref{fig:beating_n20_1m_pic}, which is narrow; \comment{however, the beating happens to be wide  in the case of BLE\,2M as there is no distinct envelope seen in Fig.~\ref{fig:beating_n20_2m_pic}}. Thus, with its fast transmission rate, BLE\,2M can fit a transmission in constructive beating crest~\cite{baddeley2023understanding}, which is not the case for BLE\,1M. 

Similarly, even though coded BLE PHYs are equipped with the FEC mechanism, the contiguous bursts of error in the transmission caused by wide beating as seen in Figs.~\ref{fig:beating_n20_125K_pic}~and~\ref{fig:beating_n20_500K_pic}  and low power were high enough to cause their built-in error-correction schemes to fail for some of the transmissions. With \NAME, BLE PHYs 125K and 500K offer around 28\% and 25\% increase in PDR, respectively, thus proving that \NAME provides robustness against wide beating. Even IEEE\,802.15.4 experiences wide beating, as seen in Fig.~\ref{fig:beating_n20_ieee_pic}. Though its DSSS error correction scheme can handle most errors, the beating width exceeded the DSSS coding depth which can cause its error correction scheme to fail, resulting in an~$\sim$8\% increase in PDR.

\section{Conclusion}
\label{sec:conclusion}

In this paper we have presented \NAME, a novel error correction scheme \textit{specifically} designed to address beating-induced errors in synchronous transmissions, an issue that has long been an acute challenge within the community. Through simulation and experimentation we have demonstrated that \NAME increases reliability by capitalizing on the inherent repetitions present in STX-based protocols and using majority voting to determine the intended value of each bit and therefore reconstruct the correct packet at the receiver. Moreover, we have shown that \NAME is applicable across different beating scenarios where devices experience `narrow' or `wide' beating dependent on the underlying physical layer~\cite{baddeley2023understanding}. 
Specifically, we show that uncoded BLE PHYs exhibit improved reliability by up to $\sim$20\% on the D-Cube testbed environment, while we additionally find that \NAME improves reliability on coded PHYs by up to $\sim$28\%.
These promising results indicate that the bit-voting technique in \NAME can successfully be used to mitigate beating-induced errors in STX-based protocols. As this work has only considered a single hop, and many STX protocols are flooding based, future work should examine the effectiveness of this technique in multi-hop scenarios as well as its scalability when considering a larger number of transmitters. 
\bibliographystyle{IEEEtran}
\bibliography{refs}

\end{document}